\documentclass[twocolumn,showpacs,preprintnumbers,amsmath,amssymb]{revtex4} 
\input psfig.sty 
\def \be {\begin{equation}} 

\def \ee {\end{equation}} 
\def \bea {\begin{eqnarray}} 
\def \eea {\end{eqnarray}} 

\usepackage{graphicx}
\usepackage{dcolumn}
\usepackage{bm}
\usepackage{epsfig} 

\begin{document} 

\title{Studying light propagation in a locally homogeneous universe through an extended Dyer-Roeder approach} 
\author{J. A. S. Lima$^1$} \email{limajas@astro.iag.usp.br} 

\author{V. C. Busti$^1$} \email{vcbusti@astro.iag.usp.br} 

\author{R. C. Santos$^2$} \email{cliviars@gmail.com} 

\vskip 0.5cm \affiliation{$^1$Departamento de Astronomia, Universidade 
de S\~ao Paulo, 05508-900 S\~ao Paulo, SP, Brazil \\  $^2$Departamento de Ci\^encias Exatas e da Terra, Universidade 
Federal 
de S\~ao Paulo (UNIFESP), Diadema, 09972-270  SP, Brazil} 


\begin{abstract} 
\noindent  Light is affected by local inhomogeneities in its propagation, which may alter distances and so cosmological parameter estimation.
In the era of precision cosmology, the presence of inhomogeneities may induce systematic errors if not properly accounted. In this vein,
a new interpretation of the conventional Dyer-Roeder (DR) approach by allowing light received from distant sources to travel in regions denser than average is proposed.   It is argued that the existence of a distribution 
of small and moderate cosmic voids (or ``black regions'') implies that its matter content was redistributed to the homogeneous and clustered matter components with the former becoming denser than the cosmic average in the absence of voids.   
Phenomenologically, this means that the DR smoothness parameter (denoted here by $\alpha_E$) can be greater than unity, and, therefore, all previous analyses constraining it should be rediscussed with a free upper limit.  
Accordingly,  by performing  a statistical analysis involving 557 type Ia supernovae (SNe Ia) from Union2 compilation data 
in a flat $\Lambda$CDM model we obtain  for the extended parameter, $\alpha_E=1.26^{+0.68}_{-0.54}$ ($1\sigma$). The effects of 
$\alpha_E$ are also analyzed for generic $\Lambda$CDM models and flat XCDM cosmologies. For both models, we find that a value of $\alpha_E$ greater than unity is able to harmonize SNe Ia and cosmic microwave 
background observations thereby alleviating  the well-known tension between low and high redshift data. Finally, a simple toy model based on the existence of cosmic voids is proposed in order to justify 
why $\alpha_E$ can be greater than unity as required by supernovae data.   

\end{abstract} 

\maketitle 

\section{Introduction} 
\label{intro} 

The accelerating cosmic concordance model (flat $\Lambda$CDM) is in agreement with all 
the existing observations both at the background and perturbative levels. However, while more data are 
being gathered, there is an accumulating evidence 
that a more realistic description beyond the ``precision era"  requires a 
better comprehension of systematic effects in order to have the desirable accuracy.   

Local inhomogeneities  are not only possible sources of different systematics, but  may also signal for an intrinsic incompleteness of the cosmic description. This occurs  because the Universe is homogeneous and 
isotropic only on large scales ($\gtrsim 100 \, Mpc$). However, on smaller scales, a variety of structures involving galaxies, clusters, and superclusters of galaxies  are observed. Permeating these structures there are also 
voids or ``black regions" (as dubbed long ago by Zel'dovich \cite{Z1}) where galaxies are almost or totally absent as recently suggested by the N-body {\it Millenium} simulations \cite{Mill}. This means that statistically uniform 
cosmologies are only 
coarse-grained representations of what is actually present in the real Universe. As a consequence,  the description of light propagation by taking into account such richness of structures  is a challenging task to improve the 
cosmic concordance model, but the correct method still remains far from a consensus \cite{rasanen2009,quartin,bolejkodr2011,Kolb2011,Clarkson1,bl12}. 

Zel'dovich \cite{Ze64}, Bertotti \cite{bert66}, Gunn \cite{gunn67}, and Kantowski 
\cite{Kant69} were the first to investigate the influence of small-scale inhomogeneities in the light 
propagation from distant sources. Later on, Dyer and Roeder (DR) \cite{Dy72} assumed explicitly that only a fraction of the average matter density must affect the light propagation in the intergalactic medium.  Phenomenologically, the unknown  physical 
conditions along the path, associated with the clumpiness effects, were   
described by the smoothness  parameter:

\begin{equation} 
\alpha = \frac {\rho_{h}}{ \rho_{h} + \rho_{cl}}, 
\end{equation} 
where $\rho_{h}$ and $\rho_{cl}$ are the fractions of homogeneous and clumped densities, respectively. This parameter quantifies the fraction of homogeneously distributed matter within a  given light cone. For $\alpha=0$ (empty beam), all matter is clumped 
while for $\alpha=1$ the fully homogeneous case is recovered, and   
for a partial clumpiness the smoothness parameter is restricted over the interval $[0,1]$. The reader should keep in mind that such a restriction clearly excludes the possibility 
of light rays traveling in regions denser than average. In principle, it should be very interesting to see how  the presence of cosmic voids - a key entity  nowadays - could  be considered in the above prescription.   

More recently, many studies concerning the light propagation and its effects on the derived distances have been performed \cite{Mattsson10,Rasanen2010,bolejkodr2011,Clarkson1}. 
Current constraints on the smoothness parameter are still weak  \cite{BSL2012,AL04,SL07,hzdata},  however, it is intriguing that the quoted analyses had their best fits for $\alpha$ equal to unity which corresponds to a 
perfectly  $\Lambda$CDM homogeneous model at all scales \cite{BSL2012,AL04}. More recently, some authors have also argued  for a crucial deficiency of the DR approach, and, as such, it should be replaced by a more detailed description, probably, based on the weak lensing approach \cite{Rasanen2010,bolejkodr2011}.   

In this paper we advocate a slightly different but complementary point of view. It will be assumed that the DR approach is a useful tool in the sense that it provides the simplest one-parametric description of the effects 
caused by local inhomogeneities, but its initial conception needs to be somewhat extended. This is done in two steps: (i) by allowing $\alpha$ (here denoted by $\alpha_E$) to be 
greater than unity in the statistical data analyses, and (ii) by interpreting the obtained results 
in terms of the existence of a distribution of cosmic voids or ``black regions" in the Universe (see Sec. V). As we shall see,  by performing  a statistical analysis  involving 557 SNe Ia from the Union2 
compilation data \cite{Union2}, we obtain  $\alpha_E=1.26^{+0.68}_{-0.54}$ ($1\sigma$) for  a flat $\Lambda$CDM model. This $1\sigma$ confidence region shows that $\alpha_E >1$ has a very significant probability.  We also show 
that $\alpha_E$ greater than unity is also able to harmonize the low redshift (SNe Ia) and baryon acoustic oscillations (BAO) data with the observations from cosmic microwave background (CMB).

\begin{figure*} 
\centerline{\epsfig{figure=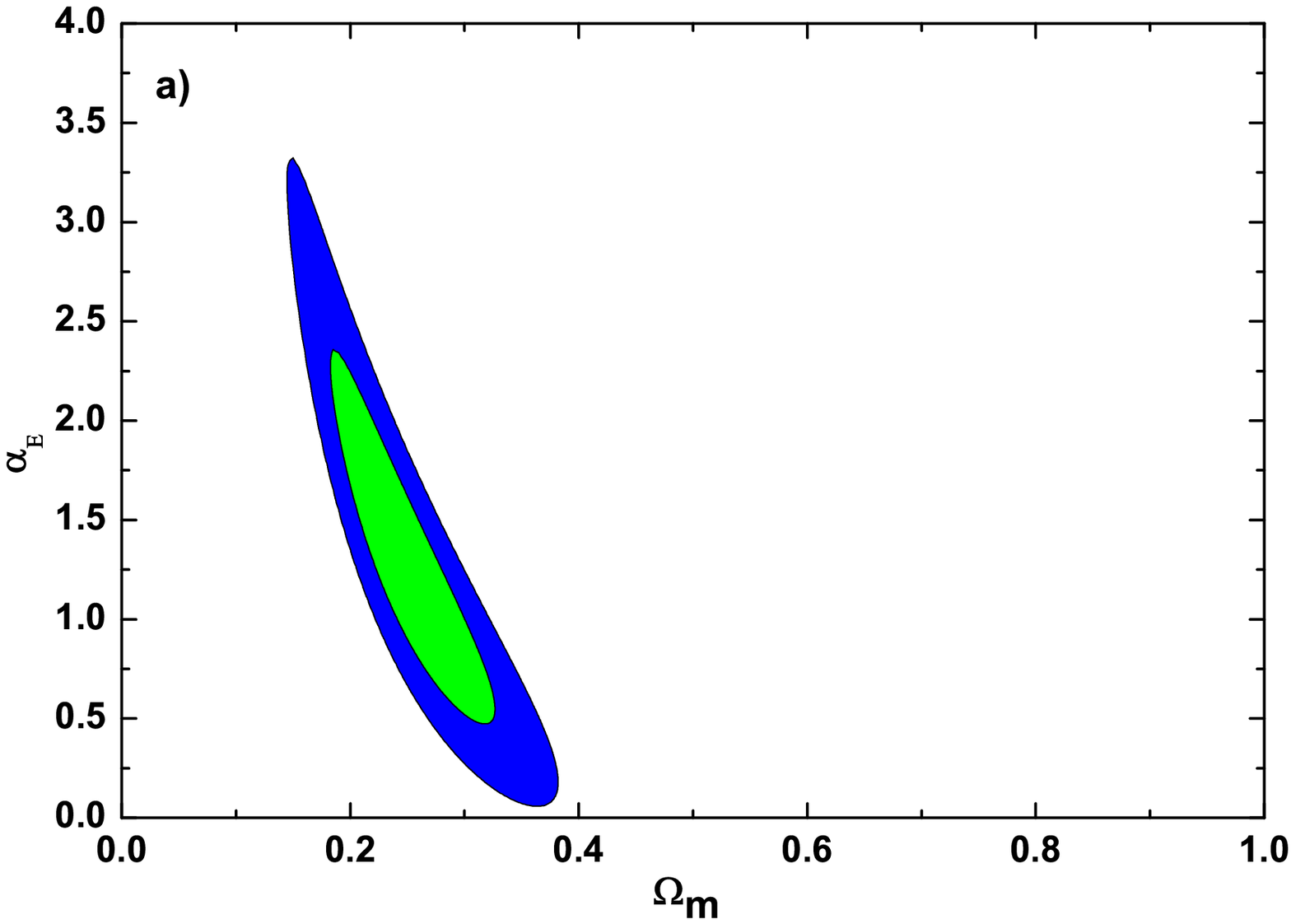,width=2.4truein,height=2.2truein} 
\epsfig{figure=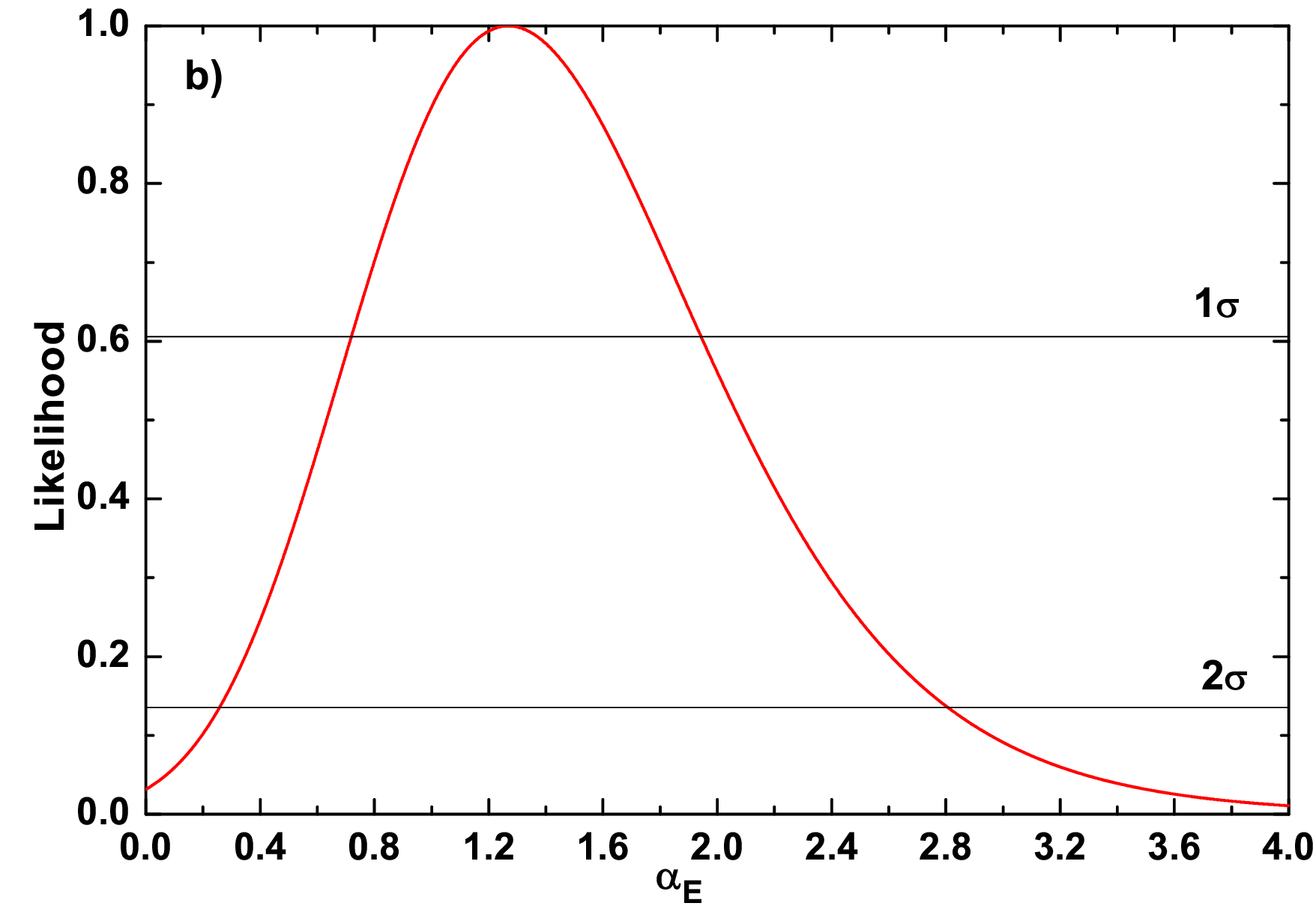,width=2.4truein,height=2.2truein} 
\epsfig{figure=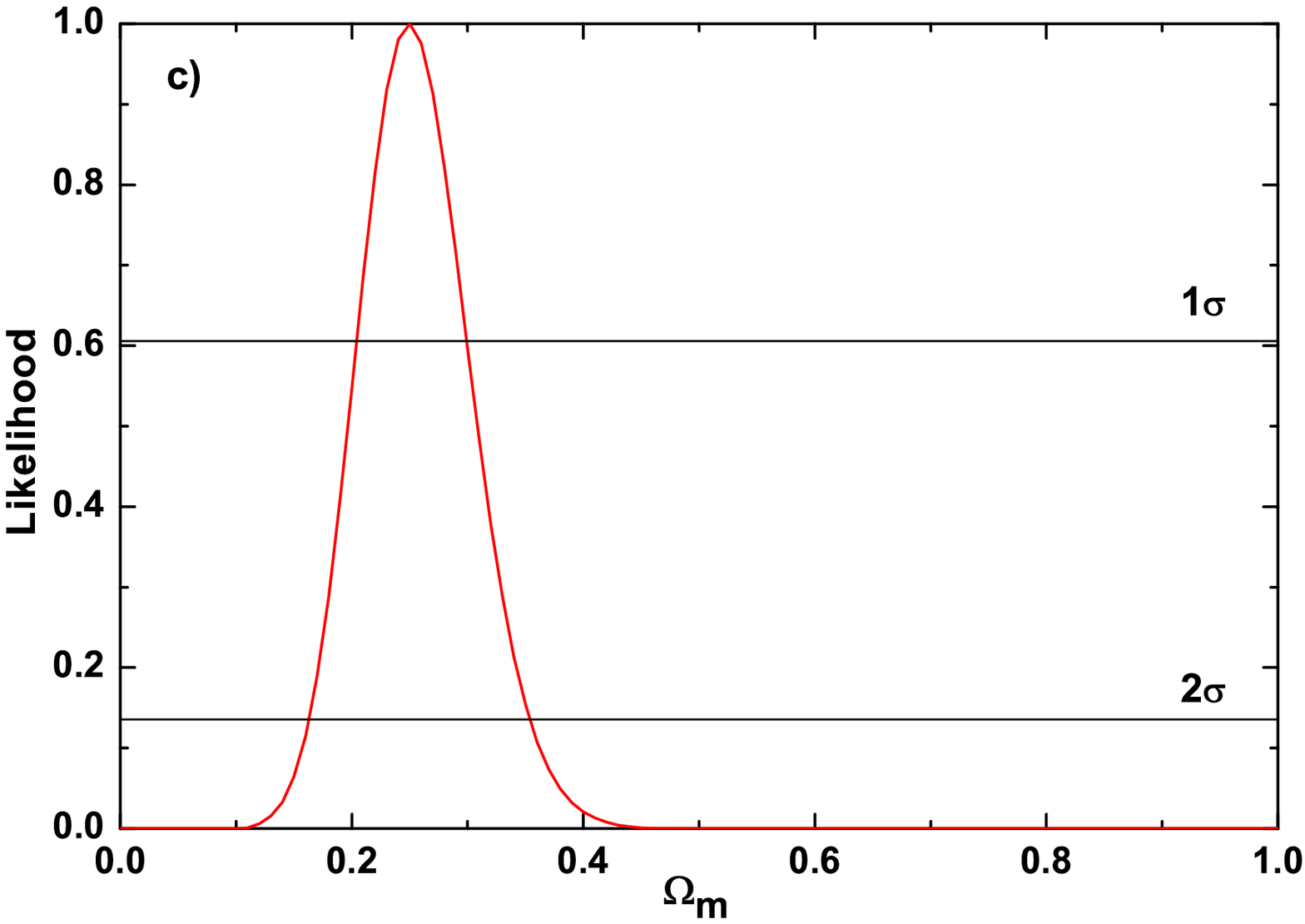,width=2.4truein,height=2.2truein}\hskip 
0.1in} \caption{(color online). {\bf{a)}} The $(\Omega_m,\alpha_E)$ plane for a flat $\Lambda$CDM model. The contours represent the 68.3\% and 95.4\% confidence levels. The best fit is $\Omega_m=0.25$ and $\alpha_E=1.26$. 
Note that a flat model with only matter and inhomogeneities ($\Omega_m=1$) is ruled out with great statistical confidence. {\bf{b)}} Likelihood of $\alpha_E$. The smoothness parameter is restricted on the 
interval $0.72 \leq \alpha_E \leq 1.94$ (1$\sigma$). {\bf{c)}} Likelihood of $\Omega_m$. We see that  the density parameter $\Omega_m$ is restricted to the interval   
$0.21 \leq \Omega_m \leq 0.29$ (1$\sigma$).} \label{figstat} 
\end{figure*} 

\section{The Dyer-Roeder Distance} 
\label{drd} 

The differential equation driving the light propagation in curved spacetimes is the Sachs optical equation 

\begin{eqnarray}\label{sachs} 
{\sqrt{A}}'' +\frac{1}{2}R_{\mu \nu}k^{\mu}k^{\nu} \sqrt{A}=0, 
\end{eqnarray} 
where a prime denotes differentiation with respect to the affine parameter 
$\lambda$, $A$ is the cross-sectional area of the light beam, 
$R_{\mu\nu}$ the Ricci tensor, $k^{\mu}$ the photon four-momentum ($k^\mu k_{\mu} =0$), and the shear was neglected \cite{Sachs61}. 

Five steps are needed to achieve the luminosity distance in the 
Dyer-Roeder approach: 

\begin{itemize} 
\item the assumption that the angular diameter 
distance $d_A \propto \sqrt{A}$; 

\item the relation between the Ricci 
tensor and the energy-momentum tensor $T_{\mu\nu}$ through Eintein's field 
equations 

\begin{equation} 
R_{\mu\nu}-\frac{1}{2} R g_{\mu\nu} = 8\pi G T_{\mu\nu}, 
\end{equation} 
where in our units $c=1$, $R$ is the scalar curvature, $g_{\mu\nu}$ is the metric 
describing a Friedmann-Robertson-Walker geometry, $G$ is Newton's constant, and  $R_{\mu\nu} k^\mu k^\nu = 
8\pi G T_{\mu\nu} k^\mu k^\nu$; 

\item the relation between the affine parameter $\lambda$ 
and the redshift $z$ 
\be 
\frac{dz}{d\lambda}=(1+z)^2 \frac{H(z)}{H_0}, 
\ee 
where $H(z)$ is the Hubble parameter whose present-day  value,  $H_0$, is the Hubble's constant;
\item the ansatz $\rho_m$ goes to $\alpha 
\rho_m$, since light experiences the local gravitational field, the matter density felt by a light ray will be $\alpha \rho_m$ instead of 
$\rho_m$, which is assumed to be valid in the line of sight of a typical light ray; and, finally, 

\item the validity of the duality relation between the 
angular diameter and luminosity distances 
\cite{ETHER33,Holanda10,Ellis2012} 
\be 
d_L(z)=(1+z)^2 d_A(z). 
\ee 
\end{itemize} 

For a general XCDM model, where the dark energy component is described by a perfect fluid with equation of state $p_X=w \rho_X$ ($w$ constant), the Dyer-Roeder distance 
$(d_L=H_0^{-1} D_L)$ can be written as

\begin{eqnarray} 
\frac{3}{2}  \left[ \alpha_{E}(z)\Omega_m (1+z)^3 +  \Omega_X (1+w)(1+z)^{3(1+w)}  \right]  D_L(z) \nonumber 
\\ + (1+z)^2  E(z) \frac{d}{dz} \left[ (1+z)^2 E(z) \frac{d}{dz} \frac{D_L(z)}{(1+z)^2}  \right]   =  0,   
\label{angdiamalpha} 
\end{eqnarray} 
where $\alpha_{E}(z)$  denotes the extended Dyer-Roeder parameter,  $\Omega_X$, $w$, are the density and equation of state parameters of dark energy while  the dimensionless Hubble 
parameter, $E(z)= H/H_0$, reads: 
\begin{equation} 
E(z)= \sqrt{\Omega_m (1+z)^3 + \Omega_X (1+z)^{3(1+w)} + \Omega_k(1+z)^2}, 
\end{equation} 
where $\Omega_k=(1-\Omega_m - \Omega_X)$ and the limiting case ($\omega =-1, \, \Omega_X = \Omega_{\Lambda}$) of all the above expressions describe an arbitrary  $\Lambda$CDM model. The above Eq.(\ref{angdiamalpha}) must be solved 
with two initial conditions, namely, $D_L\left(z=0\right) =0$ and $\frac{dD_L}{dz}|_{z=0}=1$. As in the original DR approach, from now on it will be assumed that $\alpha_E$ is a constant parameter (see, however, \cite{Linder88,SL07}). 

\section{Determining $\alpha_E$ from supernova data} 
\label{stat} 

In order to show the physical interest of the approach proposed here we have performed a statistical analysis involving 557 SNe Ia from the Union2 compilation 
data \cite{Union2}. Following standard lines, we have applied  the maximum likelihood estimator (we refer the reader to Ref. \cite{Union2,BSL2012} for details on statistical analysis involving  Supernovae data).

In Fig. \ref{figstat}(a) we display the results obtained by assuming a flat $\Lambda$CDM model. 
The contours correspond to 68.3\% (1$\sigma$) and 95.4\% (2$\sigma$) confidence levels. 
The best fits are $\Omega_m=0.25$ and $\alpha_E=1.26$. As we can see from Figs. \ref{figstat}(b) and \ref{figstat}(c) the matter density parameter is well constrained, being restricted over the interval $0.21 \leq  \Omega_m \leq 0.29$ (1$\sigma$), 
while the smoothness parameter is in the interval $0.72 \leq \alpha_E \leq 1.94$ (1$\sigma$). Although $\alpha_E$ is poorly constrained, we see that the probability peaks in $\alpha_E>1$, and, therefore, denser than average 
regions in the line of 
sight are fully compatible with the data. It is interesting to compare the bounds over $\Omega_m$ with our previous analysis with the restriction $\alpha_E \leq 1.0$ \cite{BSL2012}. The interval $0.24 \leq \Omega_m \leq 0.35$ (2$\sigma$) 
was obtained. As should be expected, by dropping the restriction $\alpha_E \leq 1.0$ lesser values of $\Omega_m$ are allowed 
by data. 

\section{SNe Ia-CMB tension and $\alpha_E$} 

The tension between low and high redshift data has been reported by many authors (see, for instance, \cite{sss2009}). A numerical weak lensing approach to solve this problem was recently discussed by  
Amendola {\it et al.} \cite{quartin} based on a meatball model. Can such a tension be alleviated by our extended DR approach? 

\begin{figure*}   
\centerline{\epsfig{figure=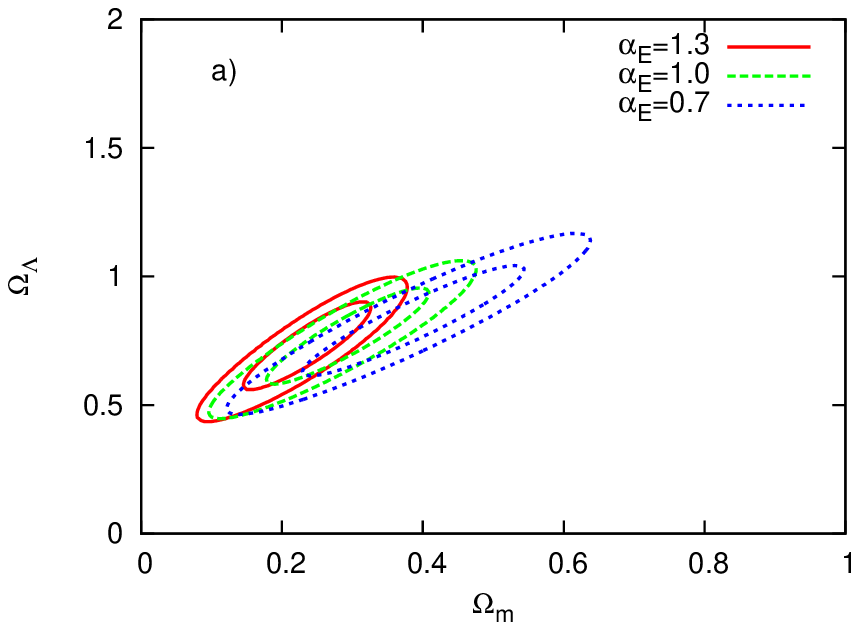,width=2.8truein,height=2.8truein,angle=-90} 
\epsfig{figure=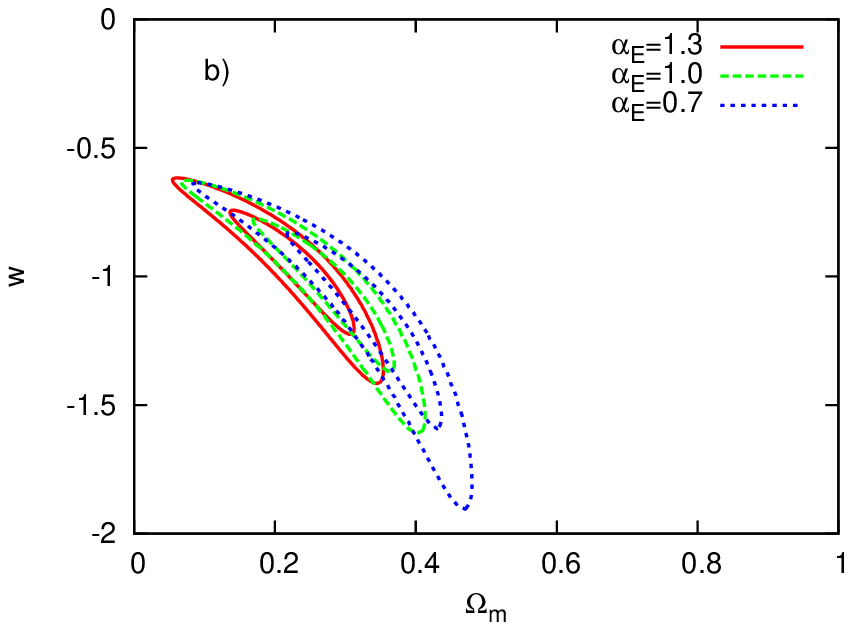,width=2.8truein,height=2.8truein,angle=-90} 
\hskip 
0.1in} \caption{(color online) {\bf{a)}} The influence of the smoothness parameter on the $(\Omega_m,\Omega_\Lambda)$ plane. The contours for three values of the 
smoothness parameter $\alpha_E$ using 557 SNe Ia from the Union2 Compilation Data \cite{Union2} correspond to 1, 2, and $3\sigma$. Greater values of $\alpha_E$ provide results more compatible with a flat model. {\bf{b)}} Contours for 
the $(\Omega_m,w)$ plane in a flat XCDM model. The same trend is observed, greater values of $\alpha_E$ imply greater values of $w$ thereby alleviating the tension among the low and high redshift data.} 
\label{figtension} 
\end{figure*}

In order to answer that, let us consider an arbitrary  $\Lambda$CDM model and plot the bounds on the ($\Omega_m,\Omega_\Lambda$) plane by  fixing three different values of $\alpha_E$. By selecting $\alpha_E=0.7$, 1.0, and 1.3 we may study what happens with the ($\Omega_m,\Omega_\Lambda$) contours when higher values are considered. 
In Fig. \ref{figtension}(a) we show the contours obtained for the chosen values of $\alpha_E$. Note that when  $\alpha_E$ grows from $0.7$ to $1.3$ the best fit moves of around $1\sigma$ towards 
lower values of the pair ($\Omega_m,\Omega_\Lambda$) thereby becoming more compatible with the cosmic concordance flat $\Lambda$CDM model. This is a remarkable  result since it improves the agreement with independent constraints 
coming from BAO and the angular power spectrum of the CMB, and, more importantly,  maintaining the same reduced $\chi^2_{red}$. 

\begin{table}[htbp] 
\caption{Best fits for $\Omega_m$ and $\Omega_\Lambda$.} 
\label{tab1}
\begin{center} 
\begin{tabular}{@{}cccc@{}} 
\hline $\alpha_E$ \hspace{0.4cm}& $\Omega_m$ \hspace{0.2cm} & $\Omega_\Lambda$ \hspace{0.4cm}& 
$\chi^2_{red}$ 
\\ \hline\hline 
0.7 \hspace{0.4cm}& 0.39 \hspace{0.2cm} & 0.83 \hspace{0.4cm}& 0.978 \\ 
1.0 \hspace{0.4cm}& 0.30 \hspace{0.2cm} & 0.78 \hspace{0.4cm}& 0.977  \\ 
1.3 \hspace{0.4cm}& 0.24 \hspace{0.2cm} & 0.74 \hspace{0.4cm}& 0.977  \\ 
\hline 
\end{tabular} 
\end{center} 
\end{table} 

In Table \ref{tab1}, the basic results are summarized. Note that the greatest value of $\alpha_E$ yields the minimum reduced $\chi^2_{red} = \chi^2/\nu$ ($\nu$ is number of d.o.f.).

In Fig. \ref{figtension}(b), we display the  statistical results for a flat XCDM model and the same values for $\alpha_E$ adopted in the previous $\Lambda$CDM analysis.  Again, we see that for higher values of $\alpha_E$, the contours 
are displaced towards regions with higher values for $w$ and smaller values for $\Omega_m$, again contributing to cancel the tension between the low and high redshift data. 

In Table II, we summarize the best fits for $\Omega_m$ and $w$ along with their respective 
minimum reduced $\chi^2_{red}$. 
\begin{table}[htbp] 
\caption{Best fits for $\Omega_m$ and $w$.} 
\label{tab2}
\begin{center} 
\begin{tabular}{@{}cccc@{}} 
\hline $\alpha_E$ \hspace{0.4cm}& $\Omega_m$ \hspace{0.2cm} & $w$ \hspace{0.4cm}& 
$\chi^2_{red}$ 
\\ \hline\hline 
0.7 \hspace{0.4cm}& 0.35 \hspace{0.2cm} & -1.18 \hspace{0.4cm}& 0.978 \\ 
1.0 \hspace{0.4cm}& 0.29 \hspace{0.2cm} & -1.06 \hspace{0.4cm}& 0.978  \\ 
1.3 \hspace{0.4cm}& 0.23 \hspace{0.2cm} & -0.96 \hspace{0.4cm}& 0.977  \\ 
\hline 
\end{tabular} 
\end{center} 
\end{table} 

\section{Why is $\alpha_E$ bigger than unity?} 

Here we propose a simple toy model based on the existence of cosmic voids in order to explain why $\alpha_E$ can be  bigger than unity. Recent studies have pointed out that cosmic voids not only represent a key constituent of the 
cosmic mass distribution, but, potentially,  may become one of  the cleanest  probes  to constrain cosmological parameters \cite{Cvoids}. The idea is to consider that very large voids are relatively rare entities, i.e. 
their formation suffers from the same kind of size (mass) segregation felt by the largest galaxies and clusters. By assuming that the three basic entities filling the observed Universe  are (i) matter homogeneously 
distributed ($\rho_h$), (ii) the clustered component ($\rho_{cl}$), and (iii) voids ($\rho_{vd}$) of small and moderate sizes, we define the extended DR parameter [see Eq.(1)]: 

\begin{equation}\label{voids} 
\alpha_E = \frac {\rho_{h}}{\rho_{h} + \rho_{cl} + \rho_{vd}}. 
\end{equation} 
The important task now is to quantify  the contribution of voids representing the local underdensities in the Universe. The presence of a void means that its matter was somehow redistributed to the clustered and the homogeneous components. The gravitational effect of a void in an initially homogeneous distribution is  equivalent to superimposing a negative density (for small densities the nonrelativistic superposition principle is approximately valid). For simplicity, it will be  assumed here that the overall  contribution of the void component can be approximated by the linear expression, $\rho_{vd} = -\delta(\rho_{h} + \rho_{cl})$, where $\delta$ is a positive number smaller than unity. Therefore,  $\alpha_E$ given Eq. (\ref{voids}) can be rewritten as

\begin{equation}\label{voids1} 
\alpha_E = \frac {\rho_{h}}{(\rho_{h} + \rho_{cl}) (1 - \delta)} \equiv \frac{\alpha}{1 - \delta}, 
\end{equation} 
which clearly satisfies the inequality $\alpha_E \geq \alpha$, where $\alpha$ is the standard DR parameter. In particular,  when the clustered component 
does not contribute we find $\alpha_E = \frac {1}{1 - \delta} \geq 1$. 

Note that negative density never happens in our approach. What really happens is that
the voids behave as effectively negative because they give mass to the other components.
The negative sign comes only to provide a means to the homogeneous part to acquire a higher density.
For example, consider we have only two components: a homogeneous part ($\rho_h$) and voids ($\rho_{vd}$). Obviously, as voids have low density,
they should give part of their mass to the homogeneous part. As a consequence, the homogeneous part will have a higher density than average.
Note that we do not have negative densities anywhere, but the voids can be treated with negative sign since they are donating their mass.

Despite being only a toy model, there is a limitation of our model which is important to stress. We consider light propagates in the homogeneous part, not crossing voids. While this can happen, it is very unlikely since voids dominate the Universe by volume. Therefore, an important step forward is to estimate in a statistical way how often light propagates in voids and how $\alpha_E$ is affected, truly incorporating the presence of voids in our approach.

The previous analyses using supernovae data implies that we have effectively  constrained the extended parameter, $\alpha_E$. How does one roughly estimate the void 
contribution from this crude model? By applying the standard DR approach to the Union2 sample, 
the best fit is $\alpha=1$, and combining with the result for a flat $\Lambda$CDM model (Sec. III), one may check that the void contribution has a best fit of $\delta \sim  0.2$. It should be important to search for a possible 
connection between the present approach and more sophisticated methods from weak lensing. 

\section{Conclusions} 
\label{conc} 

In this paper we have discussed the role played by local inhomogeneities on the light propagation  based on an extended  Dyer-Roeder approach. 
In the new interpretation light can  travel in regions denser than average, a possibility phenomenologically described by a smoothness parameter  $\alpha_E > 1$.   

In order to test such a hypothesis we have performed a statistical analysis in a flat $\Lambda$CDM model and the best fit achieved was $\alpha_E=1.26$ and $\Omega_m=0.25$, the parameters being restricted to the intervals 
 $0.72 \leq \alpha_E \leq 1.94$ and $0.21 \leq \Omega_m \leq 0.29$ within the 68.3\% confidence level. Although $\alpha_E$ is poorly constrained, the results are fully compatible with  the hypothesis of light traveling in denser 
than average regions. We have also analyzed how different values for the smoothness parameter affect the bounds over $(\Omega_m,\Omega_\Lambda)$ in an arbitrary  $\Lambda$CDM model. Interestingly, $\alpha_E>1$ improves the cosmic 
concordance model since it provides  a better agreement between low and high redshift data (Supernovae, CMB, and BAO).  The same happened when a flat XCDM model was considered with the assumption that $\alpha_E>1$. 

The obtained results reinforce the interest on  the influence of local inhomogeneities  and  may 
pave the way for a more fundamental description of light propagation in the real Universe. 

\begin{acknowledgements}
JASL is partially supported by CNPq and 
FAPESP while VCB and RCS are supported by CNPq  and INCT-Astrof\'isica, respectively. 
\end{acknowledgements}

\end{document}